\documentclass[12pt]{iopart}
\usepackage{graphicx}
\usepackage[dvips]{color}

\begin{document}

\title[Operation of graphene QHR standard in a cryogen-free table-top system]{Operation of graphene quantum Hall resistance standard in a cryogen-free table-top system}

\author{T.J.B.M. Janssen$^1$, S. Rozhko$^1$, I. Antonov$^2$, A. Tzalenchuk$^{1,2}$, J.M. Williams$^1$, Z. Melhem,$^3$ H. He$^4$, S. Lara-Avila$^4$, S. Kubatkin$^4$, R. Yakimova$^5$}

\address{$^1$National Physical Laboratory, Hampton Road, Teddington TW11 0LW, UK}
\address{$^2$Royal Holloway, University of London, Egham TW20 0EX, UK}
\address{$^3$Oxford Instruments Nanoscience, Tubney Woods, Abingdon OX13 5QX, UK}
\address{$^4$Department of Microtechnology and Nanoscience, Chalmers University of Technology, S-41296 G\"{o}teborg, Sweden}
\address{$^5$Department of Physics, Chemistry and Biology (IFM), Link\"{o}ping University, S-58183 Link\"{o}ping, Sweden}

\ead{jt.janssen@npl.co.uk}

\vspace{10pt}

\date {February 2014}

\begin{abstract}
We demonstrate quantum Hall resistance measurements with metrological accuracy in a small cryogen-free system operating at a temperature of around 3.8~K and magnetic fields below 5~T. Operating this system requires little experimental knowledge or laboratory infrastructure, thereby greatly advancing the proliferation of primary quantum standards for precision electrical metrology. This significant advance in technology has come about as a result of the unique properties of epitaxial graphene on SiC.
\end{abstract}

\maketitle

\section{Introduction}

One of the goals of the modern-day metrology is to provide quantum standards at the fingertips of the end-users, shortening the calibration chain from primary standards to the final product. A shorter calibration chain will result in a higher accuracy for end-users which can be exploited to develop more advanced test and measurement equipment and subsequently lead to societal benefits where measurement is an issue.  

Resistance metrology is one of the cornerstones of electrical metrology with most national measurements laboratories around the world providing an extensive range of calibration services across many decades of resistance value~\cite{BIPM}. The primary standard for resistance is based on the quantum Hall effect (QHE)~\cite{vonKlitzing1980} which is presently realised by a lot fewer laboratories~\cite{Jeckelmann2003}. This is because the infrastructure needed to create the QHE in conventional semiconductor systems is quite elaborate and expensive as it requires temperatures of 1~K or below and magnetic fields around 10~T. Another important barrier is the expertise needed to run a quantum Hall system and verify the correct operation and quantisation parameters. Finally, liquid Helium is becoming a scarce resource, significantly increasing in price year on year, and not readily available in every country.

A simpler, cryogen-free, system is needed if more laboratories are to realise the primary standard directly and this has recently become possible with the advent of graphene. One of the first properties observed in graphene was the QHE and it was immediately realised that it is ideal for metrology by virtue of its unique band structure \cite{Novoselov2005,Zhang2005,Novoselov2007,Giesbers2008}. The Landau level quantisation in graphene is a lot stronger than in traditional semiconductor systems which implies that both a lower magnetic field can be used and that the low temperature constraint is more relaxed~\cite{Novoselov2007}. Following the original demonstration of high-accuracy quantum Hall resistance measurements in epitaxial graphene grown on SiC~\cite{Tzalenchuk2010} and proof of the universality of the QHE between graphene and GaAs~\cite{Janssen2011b}, recently these results have been very nicely reproduced by a number of different research groups~\cite{Satrapinski2013,Kalmbach2014,Lafont2015}. Particularly, a recent publication by the LNE group has demonstrated that ppb-accuracy can be achieved over a large experimental parameter range~\cite{Lafont2015}. These results also demonstrate that devices which show extraordinary good quantum Hall effect at high magnetic field and low temperature are not necessarily optimum for low magnetic field and high temperature measurements.

Measurements of the QHE at low magnetic field are complicated by the fact that the carrier density needs to be reduced to a level well below the as-grown density of epitaxial graphene on SiC~\cite{Kopylov2010} (SiC/G). Unlike exfoliated graphene on SiO$_2$, gating of graphene on SiC is not straightforward~\cite{Tanabe2010,Lara-Avila2011,Waldmann2011}. Recently, a novel technique was demonstrated which creates a static top-gate by depositing ions via corona discharge~\cite{Lartsev2014}. This technique allows for a systematic control of the carrier density and both $n$ and $p$-type densities can be achieved on both sides of the Dirac point. Importantly this method is fully reversible and can be applied repeatably. Another issue with low carrier density graphene is the homogeneity. Under these conditions it is well known that electron-hole puddles form~\cite{Martin2008} induced by charged impurities, however, in epitaxial graphene the disorder strength can be of order 10~meV, comparable to flakes on boron-nitride~\cite{Huang2015}.

Here we demonstrate for the first time measurements of the QHE with part-per-billion (ppb)-accuracy in a small table-top cryogen-free pulse-tube system. Both the longitudinal resistivity $R_{xx}$ and the contact resistance $R_{c}$ were well within the limits set by the QHR guidelines~\cite{Delahaye2003}.  Using corona gating the carrier density was controlled such that the maximum breakdown current occurred just below the maximum magnetic field of our system. The noise sources in the system were reduced to a level such that the overall standard deviation of the measurements was comparable to those achieved for a conventional liquid $^4$He/$^3$He system. The system is extremely easy to operate (it has only one button) and can run unattended for months on end, providing a stable and primary resistance reference whenever and wherever it is needed.

\section{Device design and fabrication}

\begin{figure}
  \centering
 \includegraphics[scale=0.5]{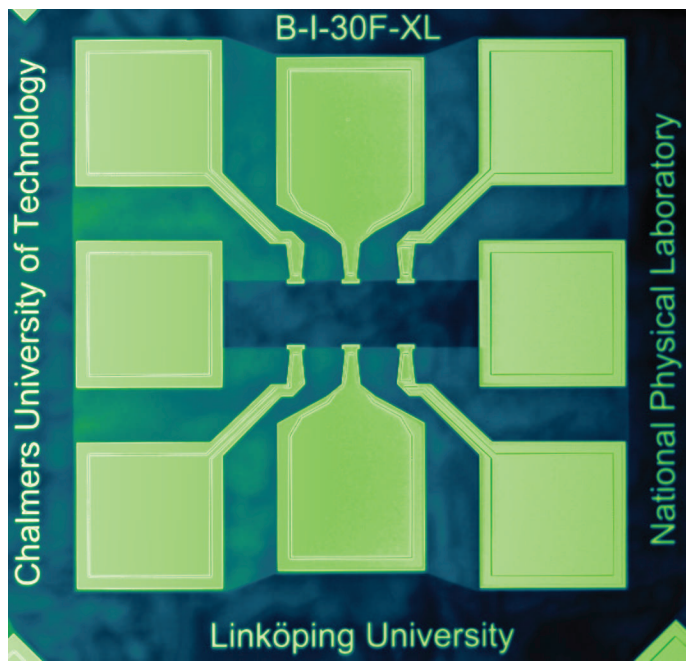}
\caption{\label{Fig1} Optical microscope image of a typical device used in our experiments (not the one used for the actual experiments). The channel width is $100\ \mu\rm m$, dark area is graphene channel, light area is SiC substrate and gold are the metallic contact.}
\end{figure}

Graphene was grown on the Si-face of SiC at $T=2000\ ^{\circ}$C and $P= 1$~atm Ar (GraphenSIC AB)~\cite{Virojanadara2008}. In total 20 Hall bars of different dimensions (30 and 100~$\mu m$ wide channels) and voltage probe types were patterned on the SiC/G using standard electron-beam lithography, lift-off, and oxygen plasma etching, as reported elsewhere~\cite{Yager2015}. The Hall bars are oriented parallel or perpendicular with respect to the predominant step edge direction of the SiC substrate. The sample was spin-coated with a thin, 100~nm, layer of poly(methyl methacrylate-co-methacrylate acid), henceforth P(MMA-MAA) (MicroChem Corp., PMMA copolymer resist solids 6\% in ethyl lactate).

All results presented in this paper are measured on a Hall bar with a 30~$\mu m$ wide and 180~$\mu m$ long channel. A comprehensive study of all devices on this chip will be presented at a later date.

\section{The measurement system}
The measurement system for primary resistance consists of two parts, the quantum Hall system and the measurement bridge.

\subsection{Table-top cryogen-free QHR cryostat}

\begin{figure}
  \centering
 \includegraphics[scale=0.7]{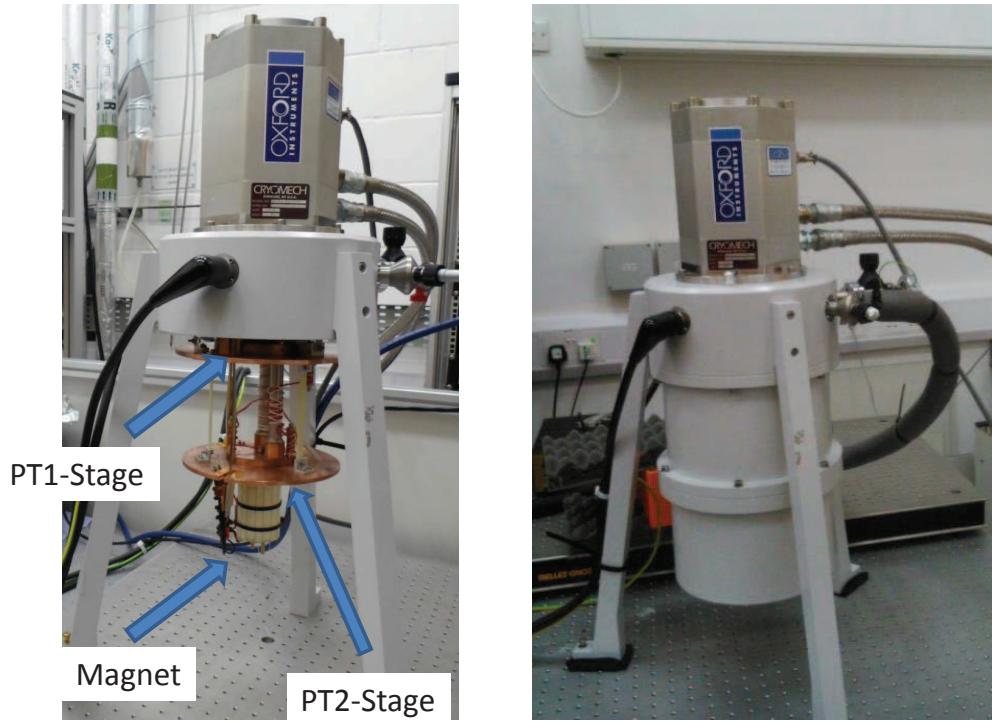}
\caption{\label{Fig2} a) Inside of the cryostat cooler showing the small superconducting magnet, mounted at the bottom of the PT2 Stage. b) The system with vacuum can mounted. The overall height of the system is around 80~cm. }
\end{figure}

Today, cryogen-free superconducting magnet systems have become omnipresent in low temperature physics laboratories because of their ease-of-use and reduced operational cost. In particular, for low magnet fields, $\leq$5~T, these systems can very small and simple. The 5~T superconducting magnet in our system is only 7.5~cm tall with an outer diameter of 6~cm. The inductance is 0.5~H and is small enough to be cooled by a 0.25~W pulse-tube cooler~(see Fig.~\ref{Fig2}). The bore of this magnet is 3~cm in diameter which large enough to take a standard TO8 header used in QHR metrology. The system has high-$\rm T_C$ current leads for the magnet which requires $\sim$60~A for full field. After evacuation the system cools down in approximately 5 hours from room temperature to $\sim 3.8$~K.

\begin{figure}
  \centering
 \includegraphics[scale=0.5]{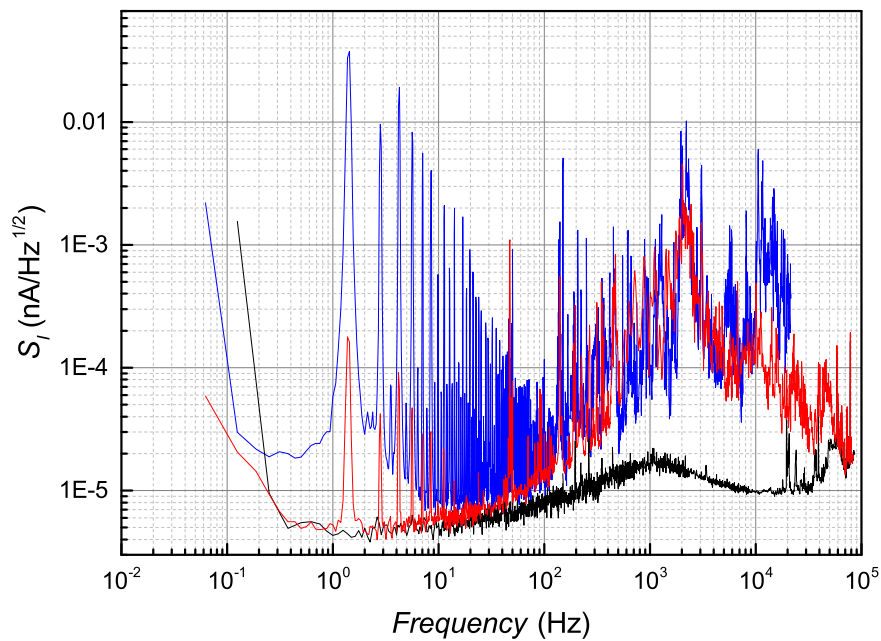}
\caption{\label{Noise} Current noise measurement traces before (blue) and after (red) modification of the pulse-tube cryostat. The black trace was measured with the pulse-tube compressor switched off. Current noise was measured on the sample wires without a sample present.}
\end{figure}

In a cryogen-free system there are a number of noise sources not normally present in a traditional wet system. There is the compressor which produces the high-pressure helium gas and the rotating valve and stepper motor on top of the cryostat. These sources of noise need to controlled and reduced as much as possible so as to not compromise the sensitivity of the measurement system. The noise of the compressor can simply be reduced by either placing an acoustic box around it or locating it in an adjoining space on the other side of a separating wall. Recently, a new type of high pressure hose was developed which significantly reduces the high-pitched hiss. These so-called quiet hoses have two beneficial effects, firstly it significantly reduces the vibrations in the cryostat system and secondly it is much more pleasant for the operator. Another improvement has been to replace the standard pulsed drive unit for the stepper motor with a low noise linear drive system. A number of other modifications are, plastic isolators on the high pressure lines to galvanically isolate the compressor from the cryostat and filters on the magnet current leads. Inside the cryostat care has to be taken that the experimental wiring is as tightly fixed as possible to reduce the effect of vibrations. Also the measurement wiring requires good heat sinking because these are relatively short compared to traditional wet systems.

The corollary of these improvements can be seen in the noise traces in Fig.~\ref{Noise}. The traces are measured on the sample wires with a spectrum analyser before and after the modifications. A reference trace with the compressor switched off is also shown. We can see that the low frequency noise peaks are reduced by more than two orders of magnitude and noise floor is equal to that measured with the compressor switched off. The higher frequency noise is largely unaffected by the modifications but this noise is outside the CCC measurement bandwidth and is not critical.

\subsection{The cryogenic current comparator bridge}
High accuracy measurements of resistance ratios are generally made using a so-called cryogenic current comparator (CCC) bridge. The fully automated CCC bridge used in our experiments has been described in great detail before~\cite{Williams2010,Janssen2012}. In a CCC bridge, currents are locked in the inverse ratio of the resistances being compared. A CCC establishes the current ratio by passing the currents along wires through a superconducting tube and measuring the residual screening current on the outside of the tube with a superconducting quantum interference device (SQUID). The difference between the voltages developed across the resistors is measured using a sensitive voltmeter and allows one resistor to be determined with respect to the other.  In primary resistance metrology one of the resistors is a quantum Hall device with a resistance value exactly equal of $R_H=R_{\rm K}/i$ where $R_{\rm K}=h/e^2$, $e$ is the elementary charge, $h$ is the Planck constant and $i$ is an integer and generally $i=2$ or $4$ is used for semiconductor devices. In graphene, owing to the bandstructure, only $i=2$ is available. The maximum achievable sensitivity of the bridge depends for a large part on the signal-to-noise ratio in the voltmeter and therefore on the maximum current used to drive the resistors (the Johnson noise in the resistors is the  other limiting factor)~\cite{Williams2010}. Under optimum conditions measurement accuracies in access of 1 part in $10^{10}$ can be achieved~\cite{Janssen2011b,Schopfer2013}. However, for routine resistance metrology a few parts in $10^9$ in a reasonable measurement time ($\sim~15$~min) is perfectly adequate. In the present system the cryogenic environment needed for the superconducting tube and SQUID is provided by a traditional liquid helium cryostat.

\section{Characterisation}

\begin{figure}
  \centering
 \includegraphics[scale=0.5]{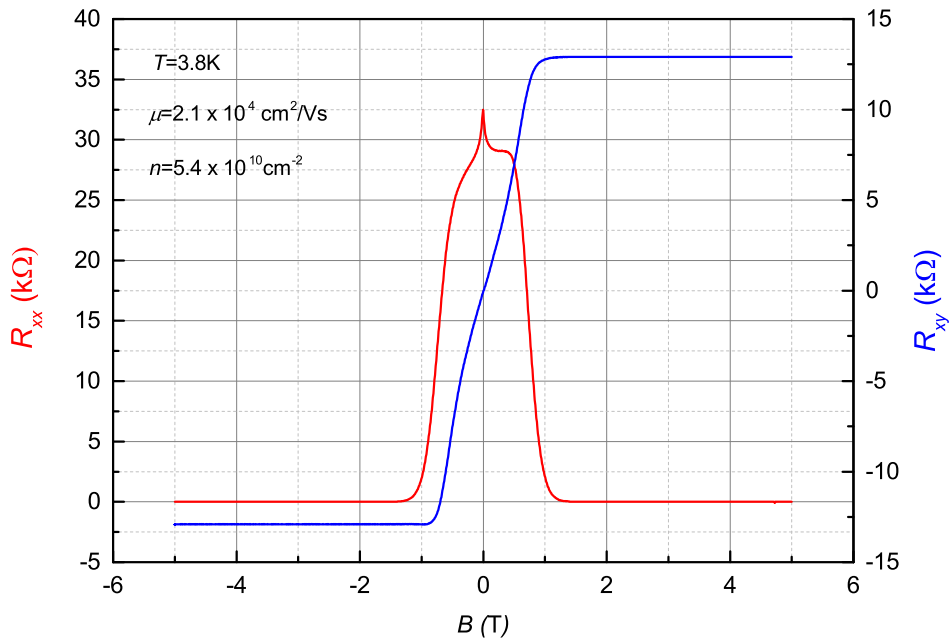}
\caption{\label{Fig3} $R_{xx}$ (Red) and $R_{xy}$ (Blue) on an epitaxial graphene Hall bar device measured at 3.8~K in the cryogen-free cryostat, measured with a source-drain current of $I_{sd}=100$~nA.}
\end{figure}

Fig.~\ref{Fig3} shows an example measurement of $R_{xx}$ and $R_{xy}$ made at the base-temperature of 3.8~K in the cryogen-free system described in the previous section. The curves display the familiar shape characteristic for epitaxial graphene on SiC which has been observed many times before~\cite{Wu2009,Shen2009,Tanabe2010,Tzalenchuk2010,Janssen2011a,Satrapinski2013,Lafont2015}. The carrier density was reduced to $5.4\times 10^{10}\rm cm^{-2}$ by corona-gating from the as grown density of $n\approx 10^{13}\rm cm^{-2}$. A wide plateau in $R_{xy}$ is observed whilst $R_{xx}$ is zero. The width of the plateau is much larger than would be expected from the low field carrier density. This behaviour is explained in terms of a magnetic field driven charge transfer from the interface layer to the graphene layer which results in an increase in carrier density as the magnetic field increases and effectively pins the Fermi level at exact filling of $\nu=2$~\cite{Kopylov2010,Janssen2011a}.

\begin{figure}
  \centering
 \includegraphics[scale=0.5]{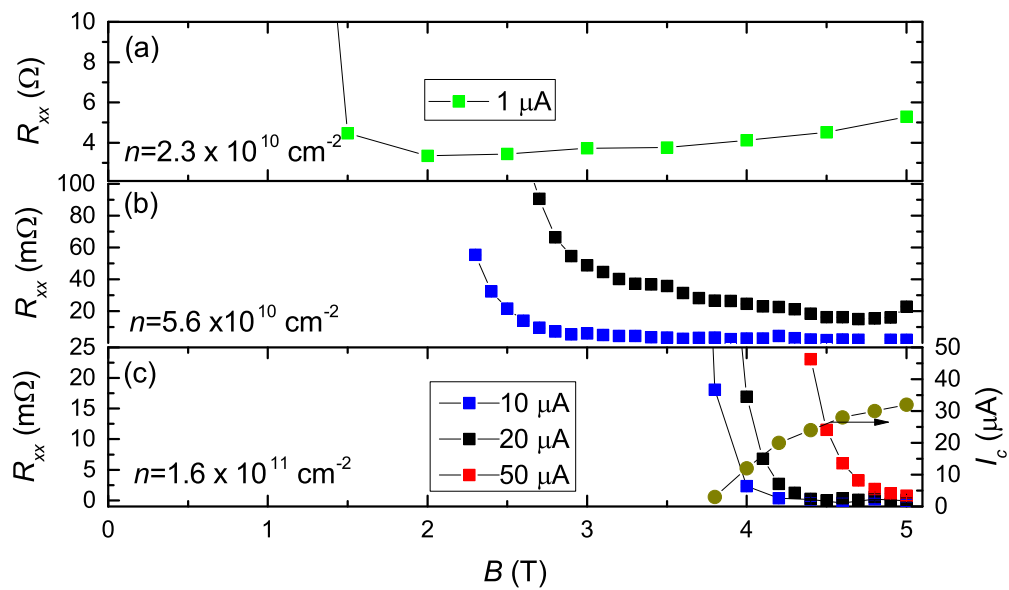}
 \includegraphics[scale=0.55]{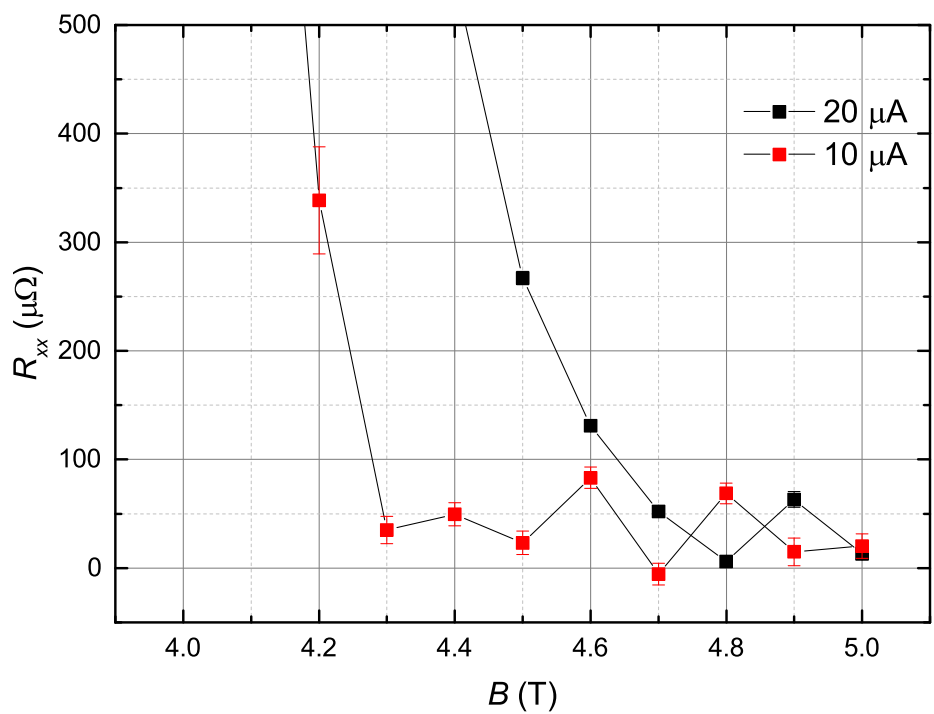}
\caption{\label{Fig4} Top graph: $R_{xx}$ as a function of magnetic field for different charge carrier densities. Temperature is $\approx$3.8~K. Plot (c) also shows the breakdown current $I_C$ as a function of magnetic field. Bottom graph: high-resolution measurement of $R_{xx}$ in a 1~T magnetic field range for $n=1.6\times 10^{11}\ \rm cm^{-2}$. This resolution was obtained by repeated measurements (typically 50 to 100) of $V_{xx}$ with positive and negative $I_{sd}$.}
\end{figure}

When attempting to make accurate quantum Hall resistance measurements the first step is to properly characterise the sample according to guidelines set out for primary resistance metrology~\cite{Delahaye2003}. Key parameters are the longitudinal resistance ($R_{xx}$) and contact resistance ($R_c$) at the desired measurement current. The longitudinal resistance needs to be as low as possible and preferably below a few tens of $\mu\Omega$ and checked on both sides of the device. Often these measurements are limited by the resolution of the nanovoltmeter and other methods can be employed to verify accurate quantisation~\cite{Delahaye2003}. The contact resistance can be accurately determined using a three terminal measurement technique in the quantised Hall state. This method determines $R_c+R_l$ where $R_c$ is the contact resistance and $R_l=6.4\ \rm\Omega$ is the lead resistance in the cryostat in our system. For our device we find $R_c$ between $0.1$ and $1\ \Omega$ for all contacts measured with a current of $10 \rm  \mu A$.

The optimum conditions for QHR measurements are easiest to obtain when the breakdown current is maximum and significantly larger than the source-drain measurement current, $I_{sd}$. Here the breakdown current is defined as the maximum source-drain current the device can sustain before a measurable longitudinal resistance appears~\footnote{We typically use a limit of 10~nV for $V_{xx}$ which for $I_{sd}=10\ \rm\mu A$ would imply a $R_{xx}=1\ \rm m\Omega$}. For higher carrier density devices, the breakdown current tends to be higher because the $\nu=2$ state occurs at a higher magnetic field~\cite{Alexander2013} which is simply related to the fact that at higher magnetic field, the Landau levels are further apart and hence the quantisation is stronger~\cite{Alexander2013}. For epitaxial graphene $I_{sd}$ was shown to follow a $\propto B^{3/2}$ behaviour similar to that observed in semiconductor systems~\cite{Jeckelmann2001}.

This effect poses a particular problem for optimising the carrier density for accurate QHR measurements at the low magnetic fields available in our small cryogen-free system. If the carrier density is too low the maximum in the breakdown current will occur at a very low magnetic field and its value will be equally low. Fig.~\ref{Fig4}a shows a measurement of $R_{xx}$ at a $n=2.3\times 10^{10}\ \rm cm^{-2}$ very close to the Dirac point. For a $I_{sd}=1\ \rm\mu A$ we find that the longitudinal resistance is always larger than a few Ohms and consequently the device is not properly quantised. Fig.~\ref{Fig4}b shows $R_{xx}$ at a $n=5.6\times 10^{10}\ \rm cm^{-2}$ and we can observe proper quantisation in a 2~T-range for $I_{sd}=10\ \rm\mu A$ but for $I_{sd}=20\ \rm\mu A$, $R_{xx}$ is in the $m\Omega$ range and the device becomes unquantised (see below). When the carrier density is set even higher (see Fig.~\ref{Fig4}c), quantisation becomes stronger but the usable magnetic field range shrinks to around 0.5~T. The bottom graph in Fig.~\ref{Fig4} shows a high-resolution measurement of $R_{xx}$ in this range demonstrating longitudinal resistance of order $10\ \mu\Omega$ and confirming proper quantisation.

\begin{figure}
  \centering
 \includegraphics[scale=0.5]{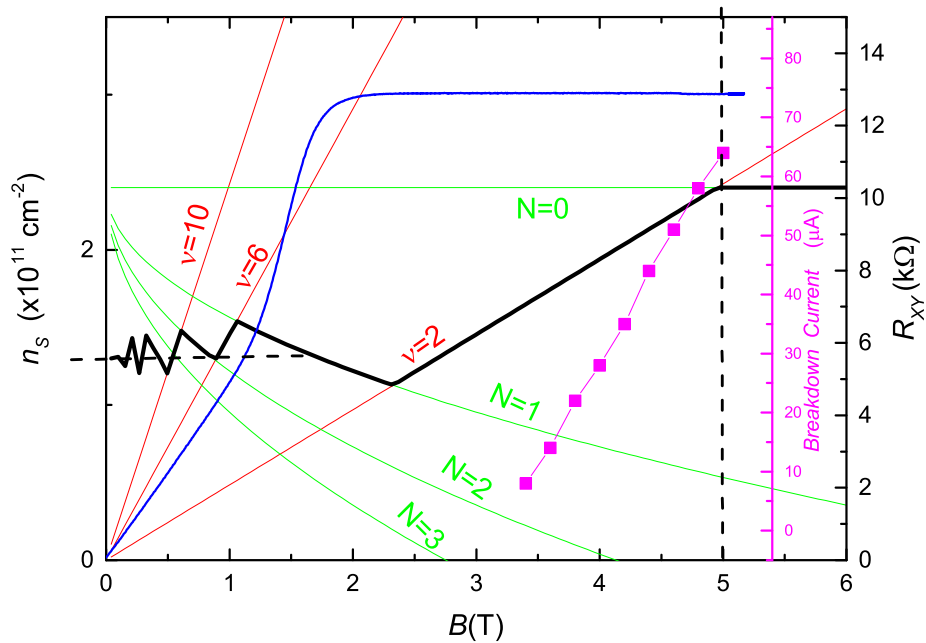}
\caption{\label{Model} $n_S$ versus magnetic field using the model from Ref.~\cite{Janssen2011a} (thick black line). Red lines are constant filling factors and green lines are $n_S(B,N)$. Blue line is $R_{XY}$ measured for a device with $n_S\approx 1.3\times 10^{11}\ \rm cm^{-2}$ (right hand axis) together with the measured breakdown current (purple squares and second purple right hand axis). Vertical dashed line indicates maximum magnetic field of 5~T and horizontal dashed line indicates zero field carrier density of $1.3\times 10^{11}\ \rm cm^{-2}$.}
\end{figure}

Using the magnetic field dependent charge-transfer model it is straightforward to estimate the optimum charge carrier density for maximum breakdown current~\cite{Janssen2011a}. Assuming that the maximum breakdown current will occur when $\nu=2$ filling factor coincides with our maximum magnetic field of 5~T~\cite{Alexander2013}, gives a carrier density of $\approx 2.4\times 10^{11}\ \rm cm^{-2}$. Setting this density as $n_{\infty}$ in the model calculation of Ref.~\cite{Janssen2011a} allows us to obtain the zero field carrier density. $n_{\infty}= \frac{A\gamma}{1+e^2\gamma/c_c}-n_g$ in which $A$ is the difference in work function between graphene and the donor states in SiC, $\gamma$ is the density of donor states, $c_c$ is the classical capacitance and $n_g$ is the deposited corona gate charge. Using $\gamma$ as a fit parameter we obtain a value for the optimum carrier density of  $n_S\approx 1.3\times 10^{11}\ \rm cm^{-2}$ (see Fig.~\ref{Model}). 

\begin{figure}
  \centering
 \includegraphics[scale=0.4]{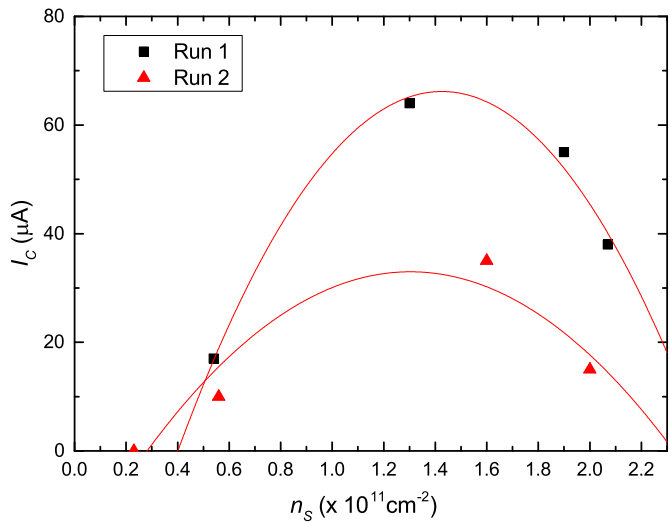}
\caption{\label{Fig4b} Breakdown current, $I_c$ versus carrier density $n_s$ at $B=5$~T and $T=3.9$~K. Black squares are for the first measurement run when the device was new and red triangles are for the second run 3 months later. Red lines are polynomial fits which serves as guide to the eye.}
\end{figure}

Figure~\ref{Fig4b} shows the measured maximum breakdown current measured at $B=5$~T as a function of zero field charge carrier density for two sets of data 3 months apart. The graph confirms that optimum carrier density is around $n_S=1.3\times 10^{11}\ \rm cm^{-2}$. For the later data set the breakdown current was almost half the original breakdown current which could be related to the degradation of one of the current contacts on the device. The cause of this degradation is yet unclear and needs to be investigated further because QHR devices for quantum resistance metrology need to be stable and reproducible over long periods of time. The original maximum breakdown current is $60\ \rm\mu A$ which for our channel width of $30\ \rm\mu m$ implies a current density of $2\ \rm Am^{-1}$ density which is close to the theoretical maximum~\cite{Alexander2013}. 

\section{Quantum Hall resistance measurements}
\begin{figure}
  \centering
 \includegraphics[scale=0.5]{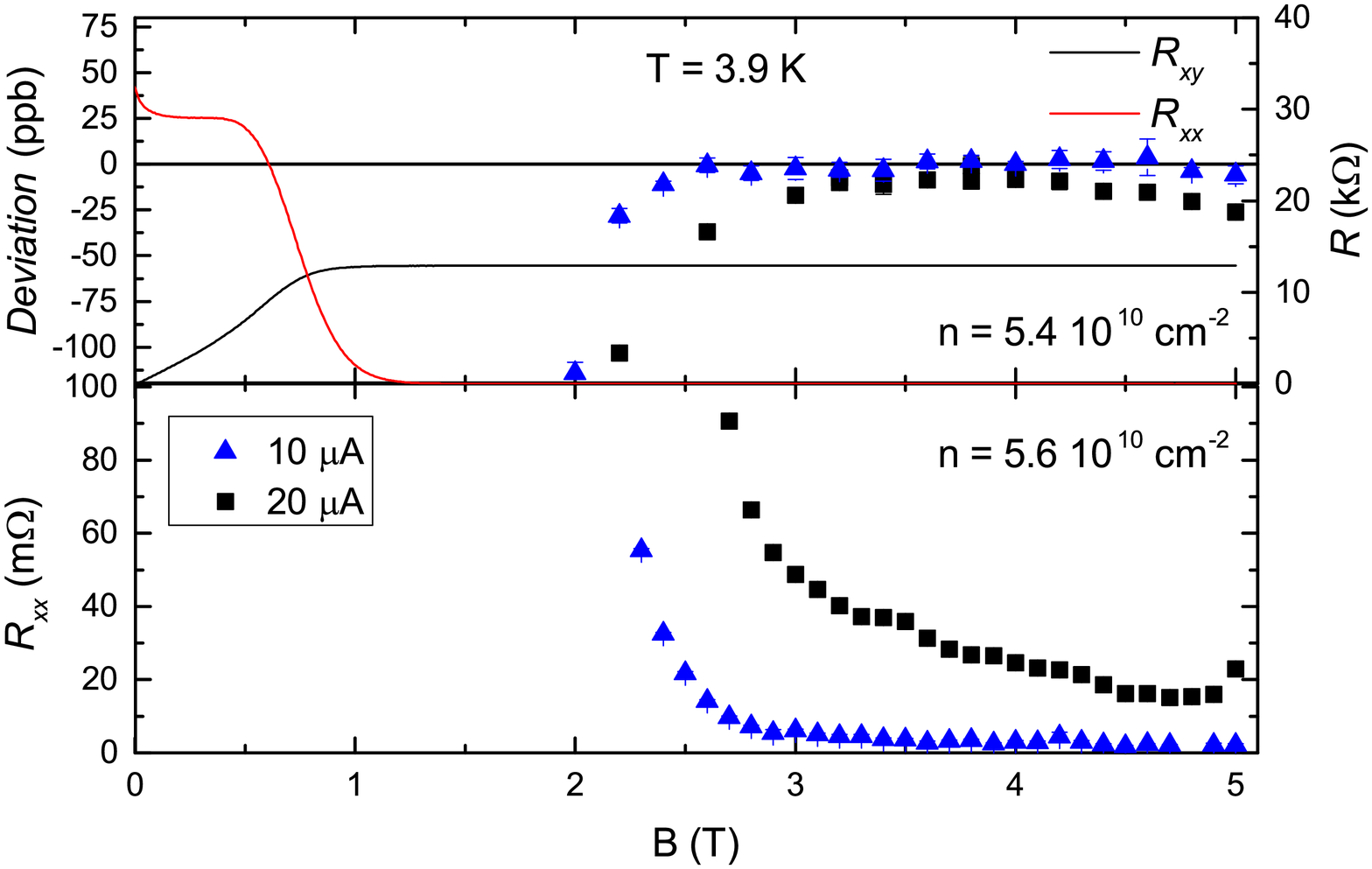}
\caption{\label{Fig6} Top panel: $R_{xy}$ (Black) and $R_{xx}$ (Red) as a function of magnetic field measured at a small (100~nA) source-drain current (left axis). Symbols: Measurement of $R_{xy}$ against standard resistor using CCC bridge. The deviation is calculated as a difference from the mean value of the standard resistor in the range of 3 to 4.5~T. Bottom panel: Measurement of $R_{xx}$ over the same magnetic field range for two different measurement currents. }
\end{figure}

Figure~\ref{Fig6} shows the central result of this paper. Here we measured the quantum Hall resistance in terms of a nominally $100\ \Omega$ temperature controlled standard resistor using the CCC bridge. The data in Fig.~\ref{Fig6} is normalised to the mean value of the resistor since we are not concerned with the absolute accuracy of the QHE in graphene which was established earlier~\cite{Janssen2011b}. The measurements are made at two different source-drain currents ($\approx$10 and $\approx$20~$\rm\mu A$) as a function of magnetic field.
Comparing the data for 10~$\rm\mu A$ with that for 20~$\rm\mu A$ it is clear that for the larger measurement current, the device is not properly quantised. This fact is also confirmed by the measurement of $R_{xx}$ which show a significant deviation from zero for this larger current.
The low breakdown current is not a major issue because the sample chip contains a number of devices with a larger width (100~$\rm\mu m$) in which the breakdown current will be correspondingly larger (to be published). For the smaller measurement current, accurate quantisation is observed over a 2~T magnetic field range which is perfectly adequate for primary resistance measurements.

The measurement resolution obtained for  most individual measurements of $R_{xy}$ in Fig.~\ref{Fig6} is 5 parts in $10^9$ for a 15 minute measuring time. A few measurements are made over a longer time (several hours) and are of order 5 part in $10^{10}$. This compares very well with traditional QHR systems, especially considering that for the cryogen-free system, there is in principle no limit on the available measurement time.

\begin{figure}
  \centering
 \includegraphics[scale=0.5]{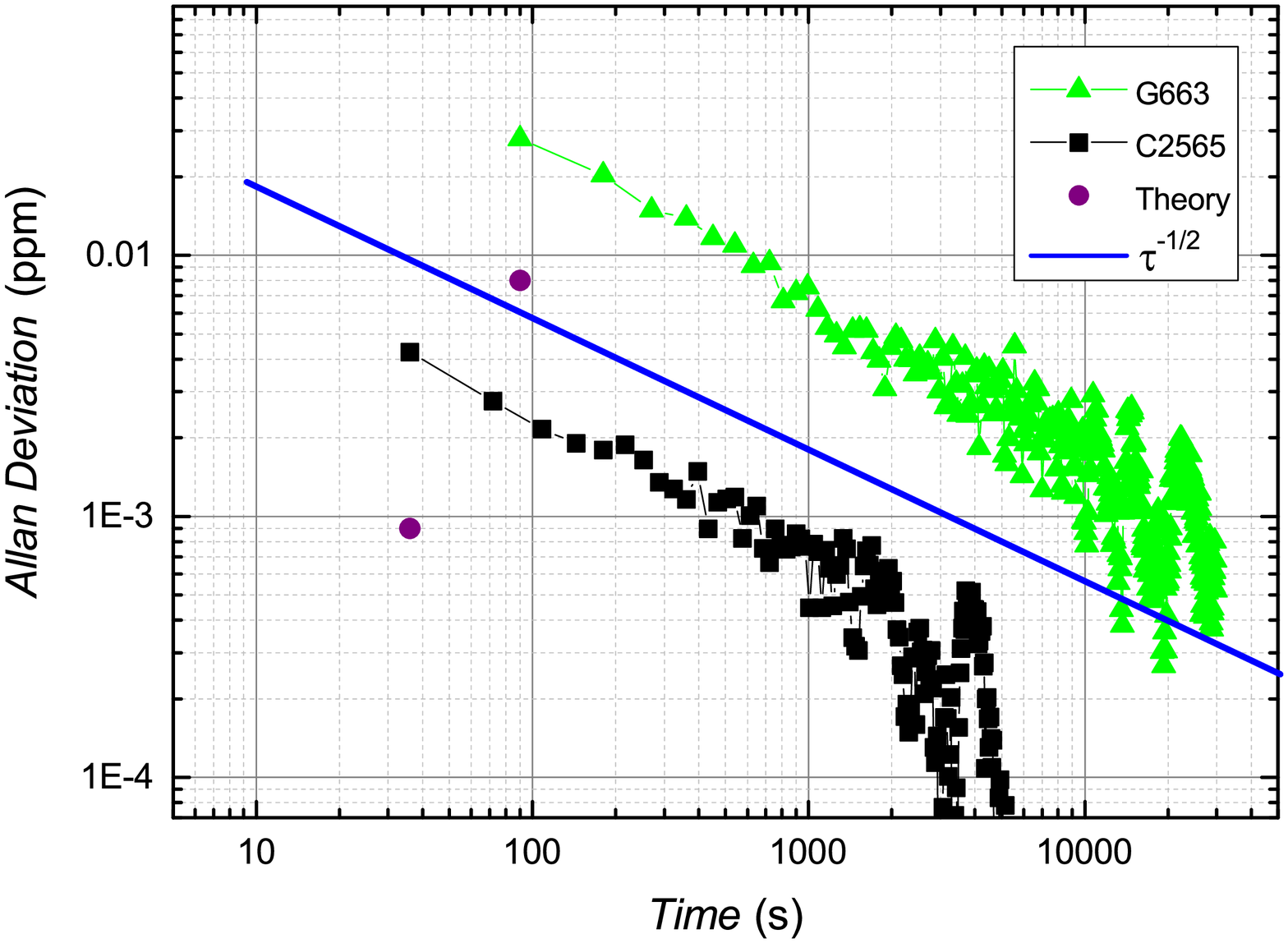}
\caption{\label{Fig5}Allan deviation for a long measurement run compared with previously published data in Ref.~\cite{Janssen2012}. Green triangles: Measured using cryogen-free system (5~T and 3.9~K) with a source-drain current of 20~$\rm\mu A$ and a CCC bridge with A20 null detector~\cite{Williams2010}. Each data point represents a 90~s measurement section composed of three 30~s measurements of either forward or reverse current direction. Black squares: Measured using traditional system with 14~T magnet-300~mK temperature and source-drain current of 100~$\rm\mu A$. The CCC bridge uses a SQUID null detector and each data point represents 30~s of measurement time made up of three blocks of 10~s. Purple dots: Theoretical optimum measurement resolution for each system. Blue line: $1/\sqrt{\tau}$. }
\end{figure}

Figure~\ref{Fig5} shows an Allan deviation plot of the measurement resolution for a long measurement run together with results obtained from a previous measurement using our standard quantum Hall system~\cite{Janssen2012}. Both curves show the expected $1/\sqrt{\tau}$ behaviour for uncorrelated white noise. The lower measurement resolution of the cryogen-free system can be explained by the lower measurement current used (20~$\rm\mu A$ versus 100~$\rm\mu A$) and the higher current noise of the null detector (A20 null-detector versus SQUID null detector), resulting in a factor of 10 difference. For both the cryogen-free system and the traditional system the theoretical optimum measurement resolution is still a factor of 5 better. This is caused by the fact that the noise of CCC-SQUID combination in our systems is about a factor of 5 higher than that of the bare SQUID sensor~\cite{Janssen2012}.

\section{Summary/Outlook}
The results presented here demonstrate that with epitaxial graphene on SiC it is possible to achieve part per billion accuracy in primary resistance metrology using a simple cryogen-free system. Measurements are presented as a function of magnetic field and different source-drain current densities which demonstrate that the operational parameters are sufficiently wide for easy and reliable use. Care has to be taken to adjust the charge carrier density to the optimum value to ensure a maximum breakdown current density. Corona gating at room temperature and subsequent freezing of the doping is beneficial compared to applying a gate voltage during QHR measurements because no additional noise is injected into the system, but this comes at the expense of the practical inconvenience of thermal cycling the system. 

Another practical aspect which needs addressing is the CCC bridge. At the moment this bridge requires a liquid helium dewar to provide the low temperature for the superconducting shield and SQUID. In a separate cryogen-free cryostat we have recently demonstrated that a CCC can be operated in such an environment (to be published). The challenge is to integrate the CCC in the same cryogen-free cryostat as the QHE system and our plan is to do this in the next design iteration of the system.

An alternative to a CCC would be a room temperature comparator bridge. In order to obtain the required ppb-accuracy a large (at least $100\ \rm\mu A$) source drain current through the quantum Hall device is needed which is beyond the breakdown current of a single SiC/G device at low magnetic field and high temperature.  In a quantum Hall array many devices can be operated in parallel, lowering the resistance value and increasing the total measurement current. The epitaxial graphene needs to be sufficiently homogeneous so that the operational parameters of all QHR devices overlap and all contacts need to be low ohmic. Recently, the first SiC/G quantum Hall array at $R_K/200$ has been demonstrated~\cite{Lartsev2015}.

Dissemination and proliferation of primary quantum standards is one of the key objectives of fundamental metrology. The results presented in this paper could be transformative for future resistance metrology by creating the opportunity for many more metrology and calibration laboratories to realise their own primary resistance traceability. This will shorten the calibration chain and lower the uncertainty which can be provided to end users with all its implicit benefits. A number of technical issues remain to be addressed but the basic principle of operation has been demonstrated.

\ack
This work was supported by the NPL Proof-of-Concept fund, NMS Programme, European Union Seventh Framework Programme
under Grant Agreement No. 604391 Graphene Flagship, and EMRP Project GraphOhm.

\subsection{References}



\end{document}